\documentclass[acmsmall]{acmart}
\AtBeginDocument{%
  }

\setcopyright{acmlicensed}
\copyrightyear{2018}
\acmYear{2018}
\acmDOI{XXXXXXX.XXXXXXX}
\acmConference[Conference acronym 'XX]{Make sure to enter the correct
  conference title from your rights confirmation email}{June 03--05,
  2018}{Woodstock, NY}
\acmISBN{978-1-4503-XXXX-X/2018/06}
\usepackage{multirow}
\begin{document}

\title{“It feels like hard work trying to talk to it”: Understanding Older Adults’ Experiences of Encountering and Repairing Conversational Breakdowns with AI Systems}

\author{Niharika Mathur}
\email{nmathur35@gatech.edu}
\affiliation{%
  \institution{Georgia Institute of Technology}
  \city{Atlanta}
  \state{Georgia}
  \country{USA}
}

\author{Tamara Zubatiy}
\email{t.zubatiy@northeastern.edu}
\affiliation{%
  \institution{Northeastern University}
  \city{Boston}
  \country{USA}
}

\author{Agata Rozga}
\email{agata@gatech.edu}
\affiliation{%
  \institution{Georgia Institute of Technology}
  \city{Atlanta}
  \state{Georgia}
  \country{USA}
}

\author{Elizabeth Mynatt}
\email{e.mynatt@northeastern.edu}
\affiliation{%
  \institution{Northeastern University}
  \city{Boston}
  \country{USA}
}

\renewcommand{\shortauthors}{Mathur et al}
\renewcommand{\shorttitle}{Conversational Breakdowns and Repair in AI Systems}

\begin{abstract}
Designing Conversational AI systems to support older adults requires more than usability and reliability, it also necessitates robustness in handling conversational breakdowns. In this study, we investigate how older adults navigate and repair such breakdowns while interacting with a voice-based AI system deployed in their homes for medication management. Through a 20-week in-home deployment with 7 older adult participant dyads, we analyzed 844 recorded interactions to identify conversational breakdowns and user-initiated repair strategies. Through findings gleaned from post-deployment interviews, we reflect on the nature of these breakdowns and older adults’ experiences of mitigating them. We identify four types of conversational breakdowns and demonstrate how older adults draw on their situated knowledge and environment to make sense of and recover from these disruptions, highlighting the cognitive effort required in doing so. Our findings emphasize the collaborative nature of interaction in human-AI contexts, and point to the need for AI systems to better align with users’ expectations for memory, their routines, and external resources in their environment. We conclude by discussing opportunities for AI systems to integrate contextual knowledge from older adults’ sociotechnical environment and to facilitate more meaningful and user-centered interactions.
\end{abstract}

\ccsdesc[100]{Human-centered computing~Empirical studies in accessibility}

\keywords{Older Adults, Conversational AI, Aging in Place, Conversational Breakdowns, Explainable AI}

\received{20 February 2007}
\received[revised]{12 March 2009}
\received[accepted]{5 June 2009}

\maketitle

\section{Introduction}
Conversational Assistants are AI systems that leverage Natural Language Processing to facilitate real-time interactions. These systems are now a part of daily life for millions of users who engage with products such as Google Home, Amazon Alexa, and Apple HomePod \cite{kepuska2018next}. The widespread adoption of these voice-enabled technologies is largely driven by their promise of convenience, hands-free operation, and seamless access to information. For many users, they assist with everyday tasks such as setting reminders, controlling smart home devices, or retrieving information quickly through voice commands. However, while these systems have become increasingly sophisticated, they still fall short in capturing the full complexity of human conversation that tends to be deeply nuanced, shaped by overlapping sociotechnical contexts, expectations, and the ability to interpret ambiguous and imprecise language. While people naturally adjust to such dynamics in human-human interactions, Conversational AI systems often struggle to do the same. When the AI’s response does not align with user expectations, misalignments or breakdowns in conversation occur. In response to these breakdowns, users employ various repair strategies, such as rephrasing commands, repeating requests, or seeking clarification \cite{beneteau2020parenting, beneteau2019communication, alghamdi2024system}. However, when these breakdowns persist, users may experience frustration or even abandon the interaction altogether, highlighting challenges for AI systems in managing the fluid and adaptive nature of dialogue and the strategies humans employ to bridge these gaps.

For older adults, these breakdowns can be particularly pronounced \cite{brewer2022if}. While voice-enabled AI assistants offer the convenience of hands-free interaction for older adults dealing with dexterity limitations, they also introduce new difficulties. Older adults may face issues such as miscalibrated expectations, conversational overlaps, and disruptions in dialog flow, making their interactions with AI systems more frustrating and unpredictable. These challenges are especially critical when AI systems are used for essential tasks, such as medication management, where reliability, trust, and communication are paramount.

In this paper, we examine conversational breakdowns between older adults and an AI assistant, focusing on how they navigate and repair these breakdowns in the context of medication management. We present findings from a 20-week deployment study involving 14 older adult participants interacting with an AI system (Google Home) designed to provide medication reminders and check-ins based on their individual schedules. Medication management offers a particularly rich context for studying conversational breakdowns, as it is inherently a collaborative process, often involving spouses, family members, or caregivers, where older adults rely on multiple sources of support, both digital and non-digital, to coordinate their routines. While collaboration is a fundamental aspect of human communication, it becomes especially significant in later life, as older adults depend on shared strategies for effective management of routines. Given the collaborative nature of both medication management and conversational repair, studying breakdowns in AI interactions provides us an opportunity to explore how older adults work with and around system limitations. By analyzing interaction data from the deployment and qualitative interviews, we investigate how AI systems can better integrate with existing social and task-based practices rather than disrupt or replace them. Our findings offer insights into how AI designers can incorporate valuable information sources into dialog flows that better align with user expectations. Furthermore, we explore older adults' envisioned expectations for future AI systems, particularly their conversational abilities and alignment with human-human dialog. These expectations not only shape how older adults engage with AI but also influence their perception of AI’s reliability and trustworthiness in supporting their needs.

To investigate these questions, we draw on sociologist Vertesi’s concept of “seams” in technological interactions, which examines how users actively navigate breakdowns by leveraging available resources in their environment \cite{vertesi2014seamful}. This framework provides a lens for understanding conversational breakdowns in AI systems, allowing us to analyze not only how older adults repair these interactions but also the human effort involved in making AI interactions successful. Rather than aiming for seamless automation, Vertesi emphasizes the value of studying the adaptive, \textit{ad hoc} strategies users employ to bridge gaps in technological interactions. Grounded in this perspective, we define our research questions as follows:

\begin{itemize}
    \item \textbf{RQ1:} How do older adults navigate and attempt to repair conversational breakdowns when interacting with a Conversational AI system designed for medication management?
    \item \textbf{RQ2:} How does the experience of encountering and repairing conversational breakdowns shape older adults' expectations for future Conversational AI systems?
\end{itemize}

The work in this paper is part of a broader effort to design assistive Conversational AI systems to support older adults aging in place. It builds on our published ASSETS work \cite{mathur2022collaborative}, where we detailed the AI system’s conversational design and shared preliminary findings from a 20-week field deployment with older adults with Mild Cognitive Impairment (MCI). MCI, an intermediate stage between expected aging and dementia, is often marked with gradual memory issues and spatial challenges  \cite{gauthier2006mild}. Given this, audio-based conversational assistants have shown promise in supporting older adults in performing routine-based everyday tasks \cite{zubatiy2021empowering, kowalski2019older}. Our initial findings in \cite{mathur2022collaborative} indicate that older adults were receptive to engaging with the AI system and found value in its ability to provide support for medication management. In this paper, we extend that work with a deeper, secondary analysis of the deployment data, focusing on conversational breakdowns and users’ \textit{ad hoc} repair strategies. We also reflect on the broader implications for designing adaptive AI systems. Figure 1 provides a chronological timeline contextualizing this study within our larger research, outlining our published prior work at ASSETS (Stage 1 – in purple) and the work presented in this paper (Stage 2 -- in pink). To reiterate, this paper is an expanded version of a conference paper published at ACM ASSETS by Mathur et al \cite{mathur2022collaborative}, focusing on secondary analysis of the interaction data, with a new set of research questions and findings, complemented by a new set of participant interviews.

\subsection{Contributions}

With this work, we provide the following three primary contributions to the larger research community:   

\begin{itemize}
    \item \textit{First}, drawing on evidence-based insights from a 20-week longitudinal study, we identify four types of conversational breakdown instances that occurred between older adults and the AI system and illustrate each with examples from interaction logs and present a data-informed categorization of these breakdowns.
    \item \textit{Second}, we identify and examine how older adults repair conversational breakdowns using \textit{ad hoc} strategies, highlighting important external information sources from their sociotechnical environment that could inform AI design in the future.
    \item \textit{Third}, through qualitative interviews, we define older adults’ expectations for future Conversational AI systems, offering insights for designing more contextually grounded and user-centered AI interactions. We also reflect on these findings in the context of advanced Generative AI systems to design more context-aware systems in the future.
\end{itemize}

\section{Background and Related Works}

In this section, we review existing works across the key threads in this research. We begin by reviewing related works on older adults’ use of Conversational AI (CA) systems. Next, we focus on medication management, reviewing existing interventions for adherence and underexplored concerns. We then transition to reviewing relevant works on examining conversational breakdowns in human-AI interactions and associated repair strategies. Finally, we discuss the sociotechnical challenges of human-AI systems, emphasizing on the one-sided expectation from users to navigate interactions. 

\subsection{Older Adults and their Use of Conversational AI Agents}

Recent scholarship in HCI has increasingly focused on how Conversational AIs can support older adults in health, well-being, and entertainment. These systems, either embedded in smartphones (Siri) or standalone devices (Alexa, Google Home), offer multimodal interactions for diverse user needs \cite{chen2021understanding, el2018towards, wargnier2016field, zubatiy2021empowering}. Morrow et al. propose a framework for using AIs for older adults for educational purposes, emphasizing dynamic functionality across cognitive changes, from initial adoption to sustained engagement \cite{morrow2021framework}. Most studies on assistive conversational technologies have focused on their general functionality, with limited attention dedicated to designing for assistive cognitive support, except for a few studies addressing cognitive behaviors \cite{koebel2021expert, wargnier2016field, zubatiy2021empowering}. These studies emphasize the importance of incorporating human-centered design to accommodate older adults' changing needs for support and position voice-user interfaces as effective tools for doing so. Furthermore, long-term deployments of Conversational AI systems have also demonstrated their potential for providing assistance in managing daily routines such as scheduling, cooking, information searching, and medication reminders \cite{duque2021automation, kim2021exploring, zubatiy2021empowering}. 

In such cases, it is imperative that the AI system should act as a partner with older adults (and their caregivers) to effectively coordinate care, while also preserving their sense of autonomy rather than imposing control \cite{sanders2019exploring, van2021spoken}. Additionally, the expectation for assistive AI systems to integrate into the broader sociotechnical environment rather than function in isolation has also been observed. This “more-than-human” perspective, described by Pradhan et al in \cite{pradhan2022more}, considers how interactions with in-home AI systems are often shaped by a host of resources within home environments. Later in this paper, we discuss how part of our findings in this work draw on and expand the idea of a distributed host of resources (both human and non-human) that older adults leverage to accomplish successful interactions with the AI system. 

\subsubsection{Anthropomorphization of Conversational Technologies by Older Adults}

Anthropomorphization, or the attribution of human traits to non-human entities \cite{epley2007seeing}, has shown to contribute to the perception of Conversational AI systems as social companions by older adults \cite{pradhan2021hey}. Voice-based interactions with AI systems often also engender socially driven responses, leading to enhanced engagement over time \cite{azevedo2018using}. Older adults’ inclination towards anthropomorphizing Conversational AI system, particularly voice-based ones, is reflected in their tendency to refer to these systems using personal pronouns \cite{jones2021reducing, liu2023older, lopatovska2018personification}. For older adults, Conversational AI assistants are frequently viewed as helpers that offload tasks, thus reducing caregiving responsibilities and providing reliable support \cite{pradhan2020use, zubatiy2021empowering}. Research shows that anthropomorphization varies across demographics, with some studies reporting younger users exhibit little to none \cite{purington2017alexa}. However, older adults are more likely to expect these systems to work as social companions than younger users \cite{corbett2021voice, oh2020differences}. 

While anthropomorphism can encourage engagement with AI systems, research has also revealed how excessive anthropomorphization can lead to unintended consequences, such as overtrust and misaligned expectations \cite{chang2023doubt, sundar2016hollywood}. Notably, studies have revealed how anthropomorphization often tends to be fluid and influenced by context, conversational style, and task severity \cite{chin2024like}. This perspective underscores the need to examine conversational interactions with AI systems within specific task contexts and accounting for factors such as the severity involved in the task. This is in order to avoid over-generalizing findings over all types contexts and enabling a deeper reflection on the influence of a task on the overall conversational experience \cite{chattaraman2019should}. In our study, we analyze interactions within the task of medication management which presents unique challenges for older adults. This task is collaborative, dynamic, and routine for older adults, making it a critical domain for examining AI interactions. The following section delves deeper into existing research on technological interventions for medication management among older adults, relevant to understanding the interaction-specific aspects in our work.

\subsection{Role of Technology in Medication Management in Older Adults}

Many research prototypes and commercial products integrate technology into medication management. However, such studies primarily highlight roadblocks related to ageist technological stereotypes, which often characterize older adults by their \textit{deficits rather than capabilities} \cite{barros2021circumspect, vines2015age}. Common interventions include smartphones \cite{suzuki2014smartphone}, alarms, reminders \cite{lee2014real}, automated pill dispensers \cite{mugisha2017framework}, and digital calendars \cite{bond1991detection}. However, these technologies pose usability challenges for older adults. Those newly diagnosed with MCI often feel distress over losing control of their schedules, fueling their resistance to adoption of these technological solutions \cite{barros2021circumspect}. Additional barriers to designing such systems include cost, onboarding, and technical complexity \cite{pater2017addressing}. Existing reminder-based systems also risk accidental over-medication, particularly among older adults with MCI \cite{beard2013making, rizk2020snooze}. While some research explores specialized medication management systems, few, if any, focus on integrating these into existing habits, such as calendars or notes, or involving older adults in the design process \cite{pradhan2025no}. This presents an opportunity to create systems that leverage older adults’ existing familiarity with CAs for a more cohesive experience.

In this context, our work strongly advocates for an assets-based design approach which effectively integrates within the existing strategies used by older adults to manage medications \cite{klee2014asset}. To address these gaps, we developed MATCHA, a Google Home-based medication management system designed to simplify setup and maintenance, align with familiar strategies, and reduce risk of over-medication. In this paper, we first discuss the preliminary investigation of the usability of MATCHA (described in detail in \cite{mathur2022collaborative}), and then utilize MATCHA as a probe to conduct a secondary analysis on data to analyze the conversational breakdowns and repair work that occurred during interactions. To better contextualize our findings related to conversational breakdowns and repair in this paper, we now review existing work that has examined conversational breakdowns in human-AI interactions. 

\subsection{Conversational Breakdowns and Repair in Human-AI Interactions}
\subsubsection{Motivation for Examining Conversational Breakdowns}

In the context of human-AI interactions, we define conversational breakdowns as instances during which the AI’s response fails to align with users’ expectations, and is unable to fulfil their conversational goals. In designing conversational technologies, there is a growing body of work examining such conversational fault lines or breakdowns between users and AI systems \cite{cowan2017can, luger2016like, sciuto2018hey}. While many of these studies address the appearance of breakdowns in human-AI interactions, less attention is given to learning from users about their experiences of encountering and repairing those breakdowns. The post-study evaluations for longitudinal studies of Conversational AI usage primarily evaluate overall user experience in for non-complicated and straightforward tasks like information searching or entertainment browsing \cite{duque2021automation, kim2021exploring, wargnier2016field, jiang2013users, lohse2008try}. While these evaluations provide valuable formative insights, they often overlook the complexities users face when encountering breakdowns for more complex and personalized tasks, an important domain of inquiry given that AI systems are being increasingly used in interpersonal contexts. 

As conversational complexity increases and older adults engage in more personalized AI-assisted tasks, system limitations in maintaining supportive and contextually appropriate dialogue begin to emerge, leading to conversational breakdowns \cite{alloatti2024tag}. Prior work indicates that recurrence of breakdowns such as misinterpretation of user intent, failure to recover errors, lack of explanations, etc. can erode trust and lead to system abandonment by older adults \cite{orlofsky2022older}. Trajkova et al. highlight difficulties with the systems’ unpredictable responses, particularly in critical scenarios such as seeking emergency help \cite{trajkova2020alexa}. Yang et al also characterize this unpredictability and opacity involved in AI interactions as one of the most pressing challenges encountered by AI designers \cite{yang2020re}. A comparative study highlights the age-related differences in how users perceive this unpredictability and breakdowns: they find that older adults tend to attribute conversational breakdowns with AI to the system’s failure to \textit{understand} them, while younger users see them as functional limitations of the system, reflecting their greater familiarity with technology \cite{tewari2022expecting}. As a result, Clark et al. emphasize incorporating \textit{dynamics of bond and trust} in AI design to transform these systems from functional tools into social companions for users \cite{clark2019makes}. This body of literature, among others, underscores the need for researchers to focus on older adults’ perspectives in experiencing and addressing conversational breakdowns with AI systems. Additionally, research must prioritize scenarios where trust and adaptability are paramount, such as emergency situations, medication adherence, and healthcare coordination, where older adults may face distinct challenges and employ unique strategies for managing breakdowns compared to lower-risk and routine tasks.

\subsubsection{Conversational Repair}
Understandably, there is a potential loss of trust that follows a conversational breakdown. Given this, post-hoc repair events become critical in examining how users deal with such breakdowns, prompting an increase in studying repair efforts following conversational breakdowns in recent years. In a scoping review, Alghamdi et al. categorize repair strategies into two types—\textit{system-initiated} (ex: reconfirming user response, etc.) and \textit{user-initiated} (ex: clarification, adjusting response, etc.)—based on who initiates the repair \cite{alghamdi2024system}. They find that integrating user-repair strategies into the design process presents more challenge due to the dynamic and subjective nature of user responses. Additionally, they also highlight a notable research gap in studying task-specific adaptive repair strategies, particularly those addressing users’ emotional reactions, such as verbal frustration or disengagement \cite{alghamdi2024system, brewer2022if}. This limited exploration of user-driven perspectives to resolving and repairing conversational breakdowns in a task-specific context serves as a grounding motivation in our work. Studies on the aftermath of breakdowns in human-AI interactions also emphasize the role of AI apologies in restoring trust. Voice assistants that acknowledge errors and offer recovery mechanisms are generally perceived more favorably than those that do not \cite{mahmood2022owning, mahmood2024gender, cuadra2021my}. 

In some studies, it was also observed that users attempted repairs in situ, even without prior training \cite{xu2023automation}, emphasizing how users are more likely to engage in proactive repairwork when the task is essential to their routine or immediate safety \cite{raudaskoski1990repair, ashktorab2019resilient, sharifheravi2020sa}. In our findings, we provide concrete examples of conversational breakdowns that occurred during a medication-based task and how older adults attempted to repair them, reflecting on their experiences in doing so. To establish a foundation for identifying such repair strategies, we reviewed existing studies on user responses to conversational misalignments and their subsequent actions, used as initial deductive codes to structure our analysis of the interaction data, described in detail in our methods.

\subsection{Sociotechnical Challenges and User Adaptations in Human-AI Interactions}

As the state of the art in AI advances and becomes more embedded in home life, there is a pressing need to design systems that are prepared to proactively handle conversational breakdowns. Foundational HCI research establishes that in interpersonal environments, any kind of technological systems (such as AI systems) function as sociotechnical entities, rather than isolated devices \cite{porcheron2018voice, pradhan2023towards}. In such complex environments, for users to establish successful interactions with AI systems, they are required to negotiate and engage with a host of distributed resources in their environments (for example, using a physical calendar to update appointment reminders).
The limitation of current AI system to handle such negotiations have often been likened to socially inept humans by users, such as a \textit{bad personal assistant} or \textit{bad professor} \cite{luger2016like}. In this context, authors Luger and Sellen reinterpret Don Norman’s classic HCI concept of the gulfs of execution and evaluation \cite{norman2013design} as the gap between users’ \textbf{expectations} and actual \textbf{experience} when interacting with AI systems. Likewise, Lucy Suchman’s concept of \textit{mutual intelligibility}, i.e., the capacity for individuals from different social groups to comprehend each other’s actions or language, positions human-machine communication, particularly during breakdowns, as inherently spontaneous, negotiable, and deeply rooted in the user’s situational context \cite{suchman1987plans}. 

However, research often finds that this negotiation is largely one-sided, with humans often expected to employ various “repair strategies” to sustain conversations with AI \cite{murad2019don}. A longitudinal study on AI communication breakdowns in domestic settings revealed that users code-switch and adapt discourse when conversations deviate from expectations \cite{beneteau2019communication}. Code switching is a conversational practice of transitioning from one form of language characteristic to another based on the context or audience, which Harrington et al find that older adults often employ while interacting with voice assistants \cite{harrington2022s}. 

While these adaptations highlight human creativity, they also expose the disproportionate expectation that users must align with AI rather than vice versa \cite{jiang2013users}. This burden is also evident as users are often found to shorten responses to improve AI recognition \cite{lohse2008try, maclachlan1988communication}, an inequity amplified for older adults facing cognitive challenges \cite{yu2020maps}. AI interactions often demand that users adapt their language, grasp system limitations, and use precise terminology, practices that impose cognitive strain on older adults managing caregiving and aging complexities \cite{chen2021understanding, dixon2022mobile, yu2020supporting}. 

To address these gap, we define our goals as follows: (1) identifying conversational breakdowns when older adults use AI for medication management, (2) unpacking their experiences of navigating these breakdowns, and (3) examining how they attempt to repair interactions using their situated understanding of both the AI system and their environment, shaping their expectations for future AI systems.

\section{Conversational Breakdowns Through the Analytical Lens of Seams}

While facilitating seamless interactions is often a technological design goal, it is evident that interactions in the real-world inevitably involve complexity, often leading to misalignments or breakdowns that require user intervention. Sociologist Janet Vertesi’s work on examining technological breakdowns, or what she terms as “seams”, underscores the importance of studying how people align or repair these seams, since it can reveal design opportunities to better meet their needs from technological systems \cite{vertesi2014seamful}. In her argument for seamful interactions, Vertesi highlights and brings attention to the active role that users play in managing these interactions, emphasizing how they creatively bridge technological gaps through unstructured and impromptu actions. By theoretically looking at our analysis from this perspective, we thus gain an understanding of how when older adults encounter a technological seam, they work creatively to patch those seams using resources in their environment that are external to the AI itself. 

In addition, applying the concept of seams and repair to our argument in this paper helps us in two key perspectives. First, we provide evidence-based insights that challenge the stereotype characterizing older adults as technology-resistant by highlighting their willingness to engage and persist in repairing AI interactions \cite{barros2021circumspect}. Second, through analysis of conversational interaction logs and user interviews with older adults, we identify the external information sources distributed in the home that could support the AI’s functionality, offering design opportunities for more human-centered, personalized and context-aware AI systems in the future.

In the next section, we establish context for this work by summarizing relevant research from our broader study, including our prior work published in \cite{mathur2022collaborative}. We find it important to include the perspective of our previous ASSETS paper to effectively contextualize the findings and results presented in this work, as they are based on and grounded in the design of the AI system for medication management, described in detail in \cite{mathur2022collaborative}. 

\section{Summarizing Research from Previous Work}

This section summarizes our prior published work informing the study and findings reported in this paper. In the course of our multi-year research effort to understand how older adults use AI systems \cite{zubatiy2021empowering, zubatiy2023don, mathur2022collaborative}, medication management routines (including alarms, reminders, and check-ins) emerged as a recurring and prioritized usage scenario, with 86\% of total interactions recorded for it over other tasks in a longitudinal usage study \cite{zubatiy2021empowering}. We were motivated to explore this context in further depth in subsequent studies, given the Interest in leveraging AI assistance for medication management.

Our prior work, published at ASSETS 2022, detailed the user research, design, and a 20-week field deployment of our medication management system \cite{mathur2022collaborative}. This paper extends that work by reexamining interaction data with a new set of research questions focused on identifying conversational breakdowns and how older adults addressed and negotiated these breakdowns by conducting additional qualitative interviews with them. Sections 4.1 through 4.3 below summarize our research context, exploratory research process and system design respectively\footnote{For a detailed account of this process, refer to our prior published work at ASSETS in \cite{mathur2022collaborative}}. In Section 5, we outline the data collection process and analysis specifically for the results and findings presented in this paper. 

\subsection{Setting the Context of Research}

This study is set within a U.S.-based clinical research program supporting older adults with MCI. Participants engage in a therapeutic lifestyle intervention alongside their family caregivers. The program takes a holistic approach to facilitate research on technological interventions to support aging in place. This paper is part of a broader effort exploring how Conversational AI can assist older adults with everyday tasks through participatory design and field studies. Within the program, older adults with MCI (“members”) and their caregivers are collectively termed a “dyad,” terminology that we use throughout this paper. The whole study unfolds in two stages, with Stage 1 activities shown in purple and Stage 2 in pink in Figure 1. While a detailed account of Stage 1 work is available in our ASSETS paper \cite{mathur2022collaborative}, we summarize it below for better context.

\begin{figure}
    \centering
    \includegraphics[width=0.85\linewidth]{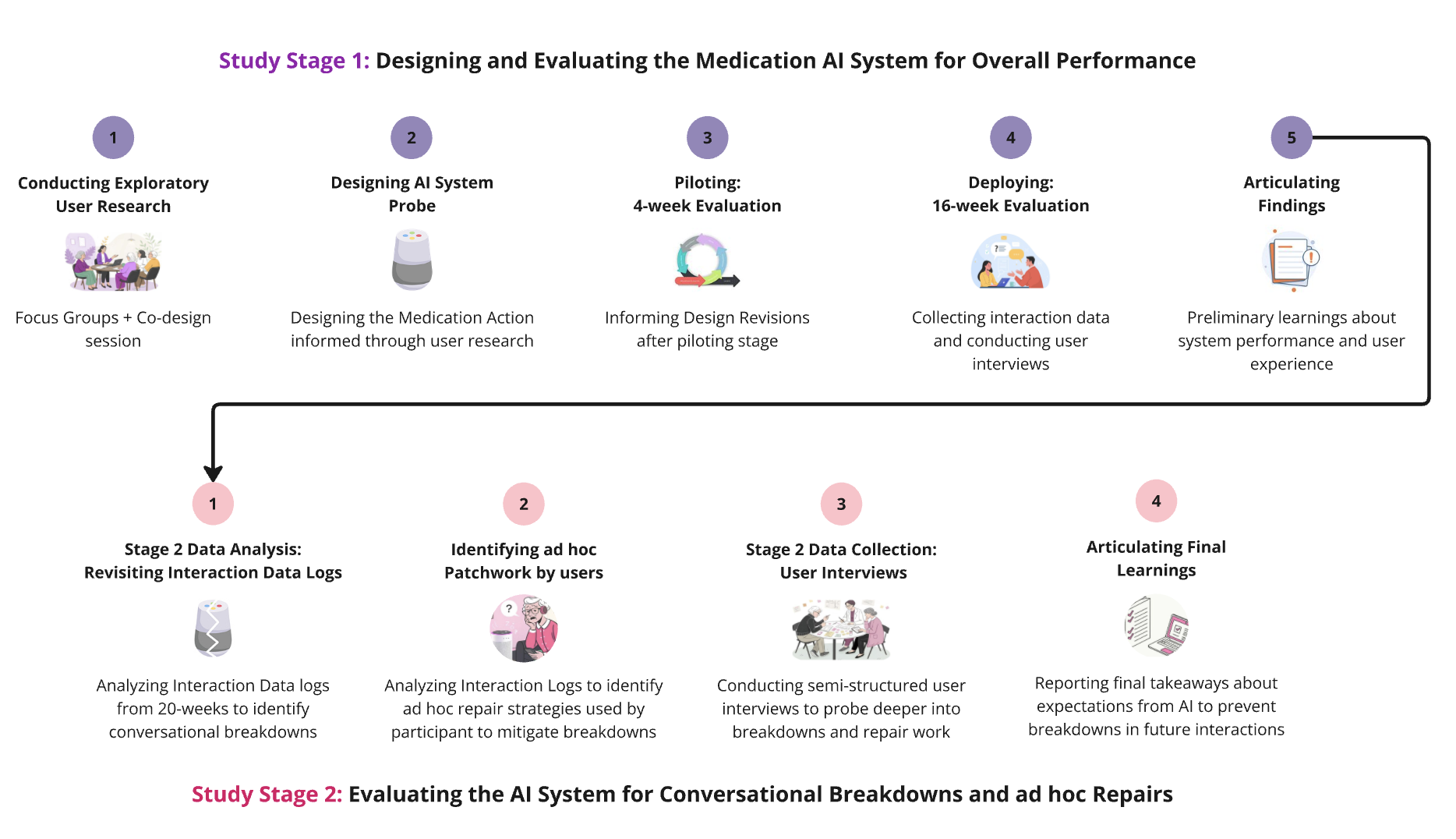}
    \caption{Two-staged structure of the study. Stage 1 (in purple) consisted of user research, system design, field deployment and preliminary evaluation (published in depth in \cite{mathur2022collaborative} and briefly summarized in this paper) . Stage 2 (in pink) consisted of additional data collection, interviews and analysis related to conversational breakdowns (reported in this paper).}
    \label{fig:enter-label}
\end{figure}

\subsection{Exploratory User Research}

In Stage 1, we conducted two focus groups and a scenario-based co-design activity to explore older adults' medication management practices (focus groups) and their expectations of AI assistance (co-design session). To ensure that AI integration complemented their existing routines rather than disrupted them, we encouraged participants to discuss their existing practices around medication management \cite{sakaguchi2021co}. The two focus groups (36 and 26 participants) lasted around 45 minutes each and were recorded, transcribed, and analyzed using affinity mapping to synthesize concrete insights \cite{harboe2015real}. The co-design activity (18 participants) used interactive storyboards of household scenarios, allowing participants to design their own AI interactions focused on the frequency of reminders and nudges. Thematic analysis of transcripts and notes informed the design goals for the medication management system, which we briefly describe in the next section.

\subsection{Conversational System Design}

Based on the design goals gleaned through the user research process, we developed the Medication Action To Check-In for Health Application (MATCHA) as a Google Conversational Action (similar to an Alexa skill). MATCHA prompts older adults to report their medication status to a Google Home assistant at pre-scheduled times set during pre-deployment interviews. The system starts with the following prompt:

\textit{"Hey John (pseudonym), it's time to take your medication. I’m checking in. Have you taken it yet?"}

After this initial prompt, the conversation then adapts based on the participant's response by using the AI's built-in, default dialog capabilities. Based on system design, each interaction can follow six possible conversational scenarios:

\begin{enumerate}
    \item (1) \textbf{"Yes"}: If the user confirms taking the medication, the AI acknowledges and ends the interaction; 
    \item (2) \textbf{"No"}: If the user requests a later reminder, the AI conversationally guides them to set a time for a future check-in;
    \item (3) \textbf{"Taking Now"\footnote{This scenario was added to the dialog flow after the initial 4-week deployment due to observed interactions, relevant reasoning for which is further described in detail in \cite{mathur2022collaborative}};}: If the user decides to take the medication immediately following the check-in, the AI acknowledges and ends the interaction; 
    \item (4) \textbf{"Don’t Remember"}: If the user is unsure about their medication status, the AI suggests checking pillbox and following up; 
    \item (5) \textbf{"Checking Pillbox"}: If the user indicates the need to check with their pillbox, the AI pauses and waits for a response;
    (6) \textbf{"Unknown"}: If the user response is outside predefined scenarios, the AI relies on its default dialog capabilities to respond. 
\end{enumerate}

\section{Method}

\subsection{Study Procedure}

We evaluated the conversational system, MATCHA, through a 20-week field study in the homes of 7 member–caregiver dyads (14 participants in total). Older adults averaged 76.4 years, and caregivers 65.4 years (participant demographics in Table 1). Participants were recruited from the clinical program, and had all engaged in the focus groups and co-design activities during the user research process. They came from diverse professional backgrounds, including technology, art, clinical support, and academia. Since they were already familiar with basic interactions with the AI, our pre-deployment training focused primarily on using MATCHA, with printed instructions and technical support contacts provided.

After pre-deployment interviews, MATCHA was installed on participants’ Google Home devices and configured to issue prompts based on individual medication schedules, identified during pre-deployment interviews. The 20-week study had two phases: an initial 4-week pilot phase followed by a 16-week phase. After the first phase, interim interviews were conducted to inform design feedback for system refinements. Participants also completed the System Usability Scale (SUS) questionnaire \cite{brooke1996sus} after the pilot phase, yielding an average score of 85.42/100, indicating general satisfaction, with some minor design revisions informed by phase 1 feedback and SUS results, described in detail in \cite{mathur2022collaborative}. 

\begin{table*}
\centering
\caption{Summary table of participant demographics.}
\begin{tabular}{|c|c|c|c|c|c|}
\hline
\textbf{\begin{tabular}[c]{@{}c@{}}Member \& \\ Carepartner ID\end{tabular}} & \textbf{\begin{tabular}[c]{@{}c@{}}Member \\ Age\end{tabular}} & \textbf{\begin{tabular}[c]{@{}c@{}}Member \\ Sex\end{tabular}} & \textbf{\begin{tabular}[c]{@{}c@{}}Daily \\ Medication \\ Frequency\end{tabular}} & \textbf{\begin{tabular}[c]{@{}c@{}}Relation to \\ member\end{tabular}} & \textbf{\begin{tabular}[c]{@{}c@{}}Caregiver \\ Age\end{tabular}} \\ \hline
M1, CP1                                                                      & 62.6                                                           & Male                                                           & 3                                                                                 & Spouse                                                                 & 60.9                                                                                                                                       \\ \hline
M2, CP2                                                                      & 78                                                             & Male                                                           & 2                                                                                 & Spouse                                                                 & 75                                                                                                                                         \\ \hline
M3, CP3                                                                      & 75.4                                                           & Male                                                           & 5                                                                                 & Spouse                                                                 & 70.6                                                                                                                                      \\ \hline
M4, CP4                                                                      & 85.3                                                           & Male                                                           & 2                                                                                 & Spouse                                                                 & 76.3                                                                                                                                       \\ \hline
M5, CP5                                                                      & 70.8                                                           & Female                                                         & 2                                                                                 & Daughter                                                               & 45.7                                                                                                                                  \\ \hline
M6, CP6                                                                      & 75.4                                                           & Male                                                           & 1                                                                                 & Spouse                                                                 & 76.7                                                                                                                                 \\ \hline
M7, CP7                                                                      & 74.3                                                           & Female                                                         & 2                                                                                 & Spouse                                                                 & 74.6                                                                                                                            \\ \hline
\end{tabular}
\end{table*}

\subsection{Data Collection and Analysis}

\subsubsection{Data Collection - Interaction data logs:} Throughout the 20-weeks, we collected interaction data logs for all participant dyads through the Google Assistant mobile app, with participant consent. These time-stamped text logs, generated through speech-to-text, captured conversational exchanges between both participants in a dyad (with voice recognition enabled to identify who was speaking) and the AI and included both AI prompts and participant responses in an .xlsx file. Two researchers cleaned the data from the "My Activity" toolbar in the Google Home app, organizing interactions by timestamps, flagging irrelevant entries (ex: pleasantries, transcription errors), and addressing any speech-to-text gaps.

\subsubsection{Data Collection - Qualitative Interviews:} We conducted semi-structured qualitative interviews at two key points during the study. The \textit{first} set of interviews were conducted immediately in the week following the end of the 20-week deployment period. The goal of these interviews was to capture overall user experience and contextualize interaction data patterns during the study period. The interview procedure for these first set of interviews, as well as the data analysis process is described in detail in \cite{mathur2022collaborative}.

The \textit{second} set of interviews were conducted at a later stage after the conclusion of the 20-week deployment period. We followed up with the participants again at this stage to conduct these additional interviews. The focus of these interviews was to probe deeper into observed conversational breakdowns, prompting them to reflect on how they navigated and attempted to resolve the breakdowns. These interviews also explored repair strategies when AI responses misaligned with user expectations and also included speculative questions on future improvements in AI systems. 

\subsubsection{Data Analysis:} We followed two distinct analysis approaches for Stage 1 and Stage 2 of our study, aligned with the research goals specific to each stage.\\

\textbf{Stage 1 data analysis:} The analysis for this stage focused on identifying interaction patterns and participant engagement during the deployment. A total of 844 interactions and post-study interviews were analyzed, and the findings from this stage, referred to as “preliminary findings” in Section 6 of this paper, helped inform the subsequent research questions and analysis in Stage 2. For a detailed description of the data collection, coding process, and findings from Stage 1, please refer to our prior publication \cite{mathur2022collaborative}. 

\textbf{Stage 2 data analysis:} \newline
\underline{Interaction data log analysis:} In Stage 2 analysis, we revisited the interaction data logs collected during the 20-week deployment, focusing on identifying conversational breakdowns and repair efforts during the interactions. Interactions are defined as a single conversational exchange between the user and the AI; the system checking-in about medication and the user providing a response counted as one interaction. In this secondary analysis, we flagged a total of 324 breakdown instances in the interaction logs. Out of these, 207 interactions (64\%) occurred in the "unknown" scenarios from Stage 1, with 117 distributed across the other conversational scenarios. Using an inductive coding approach \cite{chandra2019inductive}, two researchers collaboratively analyzed all 324 instances, generating open codes for each breakdown. In the next pass, the open codes were subsequently clustered into axial codes, which were further synthesized into high-level groups articulated as selective codes. These high-level selective codes finally informed our breakdown categories and this process resulted in the identification of four categories of conversational breakdowns (for example, all open codes pertaining to language naturalness, politeness, and speech issues were grouped into a high-level \textit{language and semantic} breakdown category).

While analyzing the interaction logs for breakdowns, we also observed instances where participants used \textit{ad hoc} repair strategies to realign conversations with the AI immediately following a breakdown. In order to have some background and structure to systematically characterize these repair attempts, we reviewed relevant literature to develop a set of \textit{deductive} repair codes. These deductive codes, drawn from the various repair strategies identified in literature are reported in the first column in Table 3. These codes provided us a starting framework for identifying repair strategies in our own interaction data, and using this deductive coding approach, we then categorized participants’ \textit{ad hoc} repair strategies across the four breakdown categories. 

\underline{Qualitative interview analysis:} Once we had analyzed the interaction logs for breakdowns and repair, we focused on the analysis of the follow-up user interviews to gain insights into participants' expectations, frustrations, and experiences in navigating those breakdowns. Two researchers transcribed these interviews and generated open codes and collaboratively refined them into themes. The resulting themes were validated across all the interviews and disagreements were resolved through merging some of the codes together, and updating the themes as needed. \\

In the following sections, we present our findings and results in the following order. \textit{First}, in Section 6, we briefly summarize the high-level findings from Stage 1, focusing on the overall perceptions of usability, engagement, and experience with the AI system. \textit{Second}, in Section 7.1, we describe the four categories of conversational breakdowns identified in the interaction logs. \textit{Third}, in Section 7.2, we discuss the repair strategies that participants employed in response to those breakdowns. \textit{Finally}, in Section 7.3, we present our qualitative findings, focusing on participant experiences of navigating and repairing breakdowns with the AI system during the study duration.

\section{Preliminary Findings from Stage 1 Analysis}

In Stage 1, we evaluated the deployment of the medication management AI system over 20 weeks. Here, we briefly summarize key insights relevant to our current analysis of conversational breakdowns and repair. Following Stage 1 evaluation, we found that MATCHA recorded a steadily rising engagement rate with it, integrated well within participants’ existing Google Home setups, increasing their confidence in maintaining their medication routines by leveraging familiar tools in their environment like their pillboxes. We also found that the usability was further enhanced by providing contextual and conversational check-ins throughout the day, rather than traditional reminders and alarms that are often repetitive and passive. The older adults in the study and their caregivers found value in the ability of the AI system to provide such personalized and gentle check-ins throughout the day, which reduced cognitive strain and supported their caregiving dynamics. With these findings, we were able to highlight the importance of designing AI systems that align with older adults’ existing routines, habits, and the sociotechnical environments around them. These findings, along with system engagement patterns and trends, are described in depth in \cite{mathur2022collaborative}.

\section{Results and Findings from Stage 2 (Conversational Breakdowns and Repairs)}

\subsection{Conversational Breakdowns}

In this section, we present the four categories of conversational breakdowns identified in our interaction logs in Stage 2 data analysis: breakdowns in \textit{language and semantics}, \textit{flow}, \textit{historical understanding}, and \textit{explanations}.

Of the total of 844 interactions recorded during the deployment period, we identified 324 (38\%) instances of breakdowns. Table 2 summarizes the four types of breakdowns, along with the percent of total breakdowns and representative examples from interaction logs. The breakdown percent for each category does not add up to 100 since some interactions were counted and labeled under more than one breakdown type. For instance, an interaction like \textit{“Why do you think he has not taken the pill yet?”} was counted both as a breakdown in \textit{language and semantics} and in \textit{explanations}. The following subsections describe each breakdown in detail. To better contextualize the understanding of a breakdown, we provide relevant participant quotes selected from the qualitative user interviews, wherever applicable \cite{cassell2004essential}. Participant quotes use “CG” for the caregiver and “M” for the member (i.e., the older adult with MCI). 

\begin{table*}
    \centering
    \caption{Summary of breakdowns, description, example and percent of breakdown}
    \begin{tabular}{| p{4.5cm} | p{5cm} | p{2.7cm} |}
    \hline 
      \centering\textbf{Breakdown Type and Description} & \centering\textbf{Example} & \textbf{Percent of total (324) interactions} \\ \hline
      \textbf{Language and Semantics} (Words and phrases spoken by the AI and its alignment with how humans talk to each other) & M2: Did I take my heart pill today? \newline AI: I’m not sure about it, try again. & 39\% (126 interactions) \\ \hline
      \textbf{Flow} (Sequential nature of conversation and expectations for turn-taking in conversation) & AI: Did you take your medications today? \newline M4: Probably, let me go check with my wife.. \newline AI: Okay, would you like to set a reminder for your medication. I can help you with that. & 21\% (68 interactions) \\ \hline
      \textbf{Historical Understanding} (Recollecting and retrieving relevant past interactions reported to it by users in older interactions) & CG2: Did [M2] take his medications in the morning? \newline AI: I don’t know how to help with that.  & 68\% (220 interactions) \\ \hline
      \textbf{Explanations} (Ability of the AI to provide context-aware explanations including external information from various sources) & AI: Did you take your medications today? \newline M1: I think I did, why are you asking me now? \newline AI: I don’t know to help with that, did you take your medications? & 43\% (139 interactions) \\ \hline
    \end{tabular}
\end{table*}

\subsubsection{\textbf{Breakdown in Language and Semantics:}} This breakdown occurred when the AI’s language, word choice, phrasing, or tone, failed to meet participants’ expectations, leading to conversational difficulties. Here, the disconnect between the AI’s responses and natural human speech emerged as the main factor for the misalignment, particularly when the AI's responses felt “robotic” or overly reliant on prototypical phrases. This misalignment frustrated users, reduced trust, and led to repeated rephrasing attempts to accommodate the AI's conversational capabilities. 

M2 articulates this disconnect, comparing it to human conversation: \textit{“I once asked if I’d taken my pills today, and she just said, ‘I don’t know how to help, try again in a different way.’ I don’t talk like that to people, and I don’t expect that from her either. I just needed a simple yes or no, but she kept asking me to say it ‘differently,’ whatever that means...”}. We also found that this expectation for AI to engage conversationally is heightened for older adults, given the interactive nature of these systems and their unfamiliarity with the AI’s inner workings and limitations.

In the context of this breakdown, we found that two key semantic factors appeared to shape perceptions of the AI’s personality: (1) metaphors participants used to describe the AI’s persona based on its language and (2) the type and sincerity of apologies during error recovery. For (1), participants frequently linked the AI’s speech to different personas. CG1 described the AI as a “nerd”, saying: \textit{“Sometimes the way she talks, all structured and mechanical, makes her sound like she’s voiced by a nerdy engineer”.} Additionally, CG2 noted tonal shifts across tasks: \textit{“There are times she sounds super helpful, almost with a ‘Midwestern politeness,’ but then when we’re not discussing medications or if I misspeak, she goes back to sounding like she’s mad at me”.} When the AI provided an apology for a conversational breakdown, participants found it as a rare but sincere attempt at empathy. As M2 noted: \textit{“I actually like that when she doesn’t understand me, she apologizes really sweetly, like she genuinely feels sorry”}. 

For this breakdown, participants often responded by simplifying their language or \textbf{“code-switching”} (what CG4 describes as \textit{learning how to speak google}). In some cases, participants responded to the breakdown with excessive \textbf{politeness} or \textbf{sarcasm}, as seen in M4’s response here: \textit{“Well, thank you so much, young lady, for trying to understand me. Would you like me to repeat it \textbf{one more time}?”}

\subsubsection{\textbf{Breakdown in Flow:}} This breakdown occurred when participants disrupted the conversation flow by asking off-topic questions, revealing the AI’s limitation in handling real-time deviations from its structured dialog flow. Even when queries were task-related, such as asking about medication status while setting a future alarm, the AI failed to integrate them, treating each interaction as isolated. While one might argue that this reflects a limitation of the system’s design rather than an inherent limitation of AI systems more broadly, our findings suggest that this distinction was largely invisible to participants. In practice, participants did not differentiate between interacting with the medication action (MATCHA) and the broader AI system, largely because both were mediated through the same voice interface. As a result, they experienced the AI as a unified entity and expected it to holistically support multiple and simultaneous interactions at once. 

This breakdown thus underscores the need for AI to multi-task, in a way that it supports parallel interactions rather than the current serial approach that we observed in the interactions. For instance, if the AI was assisting a participant with setting an alarm for medication, any related but off-script user response would be ignored, with the system rigidly continuing its original task. M3 described this frustration: \textit{“It felt like talking to a wall sometimes. I’d mention something about my pills, but it just kept asking about the time for the alarm, like it couldn’t even tell I’d already moved on..”}. CG3 further expanded on this, equating human conversation to a “partner dance”: \textit{“It’s like while we follow a basic choreography, we also adapt our movements based on each other’s cues and take turns providing support, using what we know about each other already..”}. For this breakdown, participants often responded by repeating their previous response, or by using turn-taking cues such as \textit{“ok I am done talking now..”}. 

\subsubsection{\textbf{Breakdown in Historical Understanding:}} This breakdown occurred when participants expected the AI system to recall relevant information from their past interactions with it. Interaction logs showed frequent instances where participants questioned the AI about prior medication intake, often with the goal to track their adherence over time. This need was particularly pronounced among those participants who regularly monitored their medication for personal records or for reporting to healthcare providers. M2 expressed this challenge: \textit{“Our doctor asks us about our medications, but we often don’t have a way of telling him for sure...”}. 

The AI system’s lack of memory created a significant functional gap in this context. For instance, when CG2 asked: \textit{“Hey Google, did he take his blood pressure pill this morning?”}, they were met with an error response. Similarly, M5, when prompted by the AI to report their medication status, responded: \textit{“Not yet. This week’s been something. What about yesterday though? I think I did”}, the AI, in response, generated an error response. CG4 reflected on this limitation: \textit{“While I appreciate her checking in frequently, I thought she could keep track of these things and tell me if he missed a medication this morning or yesterday”.} 

Going beyond characterizing this limitation as a functionality gap, CG4 framed this issue as one of \textit{respect} by saying: \textit{“Honestly, for me, it goes beyond her understanding me, it’s about respect. As in [the AI] remembering what I said yesterday also tells me that whoever designed her put thought into the process. Otherwise, it feels like someone deliberately made it hard to talk to her”}. This breakdown underscored the AI’s inability to recognize, recall, and retrieve personalized information. As a result, it ultimately affected its sense of companionship, confidence and personalization that participants come to expect with a system that they have been interacting with for a while in their homes. For this breakdown, participants often manually reminded the system of older interactions, pointing to household items in their environment that could enhance the AI’s memory such as their pillbox. Some participants asked for help from others in the house, while others opted to change or switch tasks entirely when the AI failed to remember their older conversations. 

\subsubsection{\textbf{Breakdown in Explanations:}} This breakdown occured when participants asked for additional context or justification in response to the AI, revealing its limitations in answering “why” questions tailored to user needs. In asking for explanations, we observed a spectrum of expectations, some participants frequently questioned the AI, especially when its reminders conflicted with their own recollection. M4 voiced this frustration: \textit{“Why are you asking me to take it now? I thought I already took it”}. Others, however, displayed high confidence in the AI, assuming its check-ins were always accurate and rarely questioned the AI. A useful metaphor for this discrepancy is imagining a hypothetical “why” button on the AI’s screen—some participants would press it often, seeking detailed explanations, while others rarely invoked it, trusting the AI’s reliability. Notably, we observed that caregivers were particularly inclined to seek explanations, as they often relied on the AI to track and verify their partner’s medication adherence. 

Participants also expressed a desire for the AI to integrate with other home devices to provide context-aware explanations. CG5 suggested: \textit{“Maybe it could check if there’s actually a pill in the box to justify the reminder, I don’t know..like a tiny camera or something?”} Similarly, CG1 highlighted the need for personalized explanations: \textit{“It would be helpful if it could start by letting him know which medicine he still needs to take, since he sometimes takes one pill but leaves the other behind in the pillbox”}. These expectations underscored the importance of context-aware explanations not only for issuing reminders but also for building confidence in the AI’s reliability in this context. For this breakdown, participants often asked the AI to provide tangible evidence to support its response. Some participants also attempted to refer to previous conversations in order to \textit{jog the AI’s memory}. 

\subsection{Repair Strategies}

\begin{table*}
    \centering
    \caption{Repair strategies drawn from literature and resulting deductive codes}
    \begin{tabular}{| p{5.7cm} | p{4.5cm} | p{3cm} |}
    \hline 
      \centering\textbf{User-Initiated repair strategies synthesized from existing literature} & \centering\textbf{Deductive codes drawn from the repair strategies identified in literature} & \textbf{Type of breakdown corresponding to the repair strategies} \\ \hline
      \multirow{4}{5.7cm}{Code switching (modifying speech) \cite{beneteau2019communication} \newline
    Over articulation/increased volume \cite{beneteau2019communication} \newline
    Hyperarticulation and Enunciation \cite{myers2018patterns} \newline
Explicitly asking for clarification \cite{baughan2023mixed, alghamdi2024system}
    \newline
    Simplifying the command \cite{baughan2023mixed} \newline Recalling details from the past \cite{myers2018patterns} \newline Information adjustment \cite{alghamdi2024system} \newline Discourse scaffold (involving other family members) \cite{beneteau2019communication} \newline Adult intervention during child-AI interaction \cite{mavrina2022alexa} \newline Correcting and asking again \cite{mavrina2022alexa} \newline Repeating the last command \cite{beneteau2019communication, kisser2022erroneous, baughan2023mixed} \newline Using a new command \cite{myers2018patterns} \newline Waiting for the system to offer reasoning \cite{mavrina2022alexa} \newline Interrupting the interaction mid-sentence \cite{kisser2022erroneous} \newline Switching to alternative input modes \cite{baughan2023mixed} \newline Modulation \cite{alghamdi2024system} \newline Emotional disengagement \cite{alghamdi2024system} \newline Expressing frustration/quitting \cite{myers2018patterns} \newline Using offensive language \cite{kisser2022erroneous}} 
    & \textbf{(1) Speech and Pronunciation Adjustments:} Code switching, over articulation/increased volume, hyperarticulation and Enunciation, Repetition \newline \textbf{(2) Clarification and Information Refinement:} Explicitly asking for clarification, Simplifying the command, adding more information, using keywords only & Breakdown in Language and Semantics \\ \cline{2-3}
       & \textbf{(3) Self-Correction and Persistence Strategies:} Correcting themselves and asking again, repeating the same command, using a new command, waiting for the system to offer reasoning \newline \textbf{(4) Interaction Control and Alternative Inputs:} Interrupting the interaction mid-sentence, switching to alternative input modes, modulation & Breakdown in Flow \\ \cline{2-3}
       & \textbf{(5) Social and Contextual Mediation:} Recalling details from the past, information adjustment, discourse scaffold, adult intervention during child-AI interaction & Breakdown in Historical Understanding \\ \cline{2-3}
       & \textbf{(6) Emotional and Expressive Responses:} Emotional disengagement, expressing frustration/quitting, using offensive language & Breakdown in Explanations \\ \hline
    \end{tabular}
\end{table*}

When these breakdowns occurred, participants often attempted \textit{ad hoc} repairs to continue the conversation. For several breakdown instances, we found that participants attempted to negotiate with the AI system in order to repair and realign the conversation. 

Notably, many of these repair efforts did not lead to successful interactions due to the AI's rigid dialog flow. However, we argue that analyzing these “failed” attempts holds scholarly value for two key reasons. First, it reveals the hidden labor users perform to repair these conversational gaps, prompting a critical evaluation of conversational technologies in advanced AI systems in the context of older adults. Second, articulating these repair attempts provides insights into older adults’ expectations for conversational context building with AI systems, thus informing design implications for future systems.

To establish context to analyze these repair strategies, we synthesized existing works on conversational repair to develop a set of deductive codes that we then used as guiding codes to analyze our interaction data for repair work using a top-down deductive approach \cite{fereday2006demonstrating}. In Table 3, we visualize this deductive approach as follows: the first column consists of 19 user-initiated repair strategies observed across existing literature on conversational repair with AI systems. The second column consists of our 6 primary deductive codes drawn from the repair strategies identified in literature, reflecting the high-level categories of repair. Finally, in the third column, we report the categorization of these repair instances corresponding to the type of conversational breakdown that the participants performed them for. Identifying these repair strategies formed the basis of our qualitative interviews to probe deeper into participants' experiences of performing those conversational repairs.

\subsection{Qualitative Insights from Stage 2 Interviews}

In this section, we present and discuss key insights gleaned from the user interviews, focusing on older adults’ dealings with conversational breakdowns and their experiences of repairing them, organized into four themes (see Table 4 for a reflection of these findings in the corresponding breakdown type). 

\textit{First}, we discuss the expectation for the AI to integrate external data beyond its internal knowledge. \textit{Second}, we examine how the AI’s limited long-term understanding influences older adults’ confidence. \textit{Third}, we unpack the invisible labor involved in managing conversational breakdowns through ad hoc strategies. \textit{Finally}, we reflect on the implications of these findings for human-AI interactions, particularly as the AI’s social and functional roles increasingly overlap in supporting older adults. Participant quotes from user interviews are included, wherever relevant, to provide contextual and interpretive depth to the insights \cite{cassell2004essential}.

\subsubsection{Need for External Memory Aids}

The need for an integrated setup underscored that much of the AI’s valuable data exists outside its internal memory, embedded in the user’s environment. Throughout the study, we observed that the home environment of older adults emerged as a complex network of resources around them. This environment included not only the AI but also household items, tools, other people, and devices, such as pillboxes, calendars (both paper and digital), sticky notes, phone alarms, and smart home systems, that collectively functioned as “extended memory aids” for older adults. We observed that participants frequently attempted to integrate these external objects and devices to support their interactions with the AI, particularly during breakdowns related to historical understanding and explanations. They also expressed a desire for these external integrations to enhance the AI’s ability to provide them with more comprehensive and context-aware reminders. CG5 said: \textit{”We have a lot of things that we use to stay on top of our individual pills…our own calendar, then the CEP calendar and the app, our regular pillbox and sometimes I even leave little sticky notes for him on the refrigerator. He often doesn’t read them but what I’m trying to say is…there’s a lot that the Google [AI] is missing in our house…we have a lot going on”}. In a similar vein, CG2 suggested a smart pillbox that could communicate with the AI, saying: \textit{“It would be helpful if we could have like a.. a smart pillbox. I know someone was talking about it at some point…like a device that looks like a pillbox, but also has the ability to talk to Google and tell her when he has or has not taken the pill out of it”}, reflecting the need for broader contextual awareness in the AI’s understanding. These perspectives underscored the importance of enabling AI systems to leverage the home environment as an extended memory, thus enhancing their ability to hold context-aware interactions.

We also noted that grounding our study in a task like medication management revealed task-specific influences on the interaction. This emphasized how the need for external integrations with the AI system becomes more pronounced in collaborative and high-importance tasks such as medication management. CG2 articulated this by describing the many resources involved in managing medications: \textit{"It’s not just me remembering about my pills and his pills. Sure, I take the center stage but I also have my pillbox and his pillbox…which are separate by the way, our calendars, phone reminders, our doctor. Even the folks at our local pharmacy, all working with us to keep us on track with our medications”}. This highlighted how medication management is a joint effort between human actors (older adults, clinicians, pharmacists) and non-human actors (pillboxes, calendars, notes), shaping the expectation that an AI in this context should also consult and collaborate across these sources of information.  

In addition, we found that beyond just identifying these external data sources (something advanced generative AI systems are increasingly designed to do), effective assistance also required structuring this data in a way that the AI can access, interpret, and accurately discern user intent. Participants emphasized that the AI should recognize \textit{why} it needs to consult a particular pillbox based on context, rather than applying a one-size-fits-all approach. As CG1 explained: \textit{“Sometimes when I ask about my pillbox, I just want to know if I’m on track with the morning meds, but other times, I’m double-checking what’s left for the evening. It would be helpful if I could tell it the difference and it would understand”}. This highlighted the importance of recognizing intent in interactions, since the actual meaning or intent behind referring to a pillbox can vary across situations and users.

\subsubsection{Looking at Personalization as a Need for Long-Term Understanding in Assistive AI Systems}

While AI’s limited personalization is often attributed to data constraints or algorithmic limitations from a technical standpoint, our interviews revealed that older adults primarily saw it as a gap in the AI’s long-term understanding of them. The AI’s inability to remember and recall past interactions at appropriate times significantly influenced their trust and confidence in its ability to truly “know” them. We found that the expectation for personalized AI interactions is especially pronounced among older adults since they often anthropomorphized the AI, emphasizing how in a home setting where the AI takes on both functional and social roles, it is perceived more as a companion than a mere tool. CG3 illustrated this by comparing the AI assistant to a workplace system: \textit{“See…if this was a system in my office…she would just need to be good at specific tasks..like help me call my boss, send emails etc…doesn’t have to know me deeply. But at home, an assistant or a companion…as you call it…should know me well, like my likes, dislikes, choices and she should pick up on what I need without having to repeat myself all the time”}. This highlighted the importance of adapting AI personalization to its context of use, distinguishing between sensitive, interpersonal settings like the home and more functional environments like the workplace.

Since the AI system in our study operated primarily through voice interactions, we found that this free-flowing “conversational” modality heightened the expectation for human-like personalization. Participants expected the AI to remember and build on prior exchanges, much like human conversations. However, its inability to recall past interactions led to perceptions that it lacked long-term understanding, respect, and empathy. Without memory, the AI felt impersonal and transactional, making some participants feel as if they were “starting over” with each interaction, rather than building a relationship. M3 reflected: \textit{“It’s like talking to someone new every day…there’s very little memory of what I said before…maybe except for my name and my medication time, which is still good...”}. Beyond recalling past interactions, participants also expected the AI to adapt over time to their evolving routines, preferences, and cognitive needs. This dynamic aspect of personalization was viewed as important for maintaining the long-term relevance of the AI in their daily lives. CG5 emphasized this: \textit{“I wouldn’t mind repeating things if it was learning and improving…like I told it at one point to remind me about medications after an hour every day…she did do that on that day but never followed up on it…”}, emphasizing that the expectation for the system to register a change in medication schedule was not registered beyond the first day, without manual intervention. 

In our analysis, the lack of long-term understanding became particularly evident in the breakdowns that occurred due to limitation in the AI’s ability to recollect relevant historical interactions, or information that has already been communicated to it in previous interactions. This finding reinforced the need to recall relevant and context-aware interactions from the past as a key design expectation for Conversational AI systems. In placing an emphasis on the relevance of information retrieval, we saw that it is important for the AI to consider the appropriateness of this recollection. While many advanced generative AI systems today can store all past interactional data, the true challenge lies in the AI's ability to discern which interactions are relevant to the current context and align those with the user's present needs and expectations. This ability thereby emphasizes the importance of intent detection in not only storing, but also effectively retrieving and utilizing interaction history at appropriate times in a meaningful way.

\begin{table*}
    \centering
    \caption{Reflection of findings in corresponding breakdown type}
    \begin{tabular}{| p{5cm} | p{5cm} |}
    \hline 
      \centering\textbf{Qualitative Learnings} & \textbf{Reflected through breakdown in} \\ \hline
      Need for external memory aids & \textit{Historical understanding, Explanations} \\ \hline
      Personalization as long-term understanding & \textit{Historical understanding} \\ \hline
      The invisible labor & \textit{Historical understanding, Flow} \\ \hline
      Speaking like a Human & \textit{Language and Semantics, Flow} \\ \hline
    \end{tabular}
\end{table*}

\subsubsection{The Invisible Labor in Mitigating Conversational Breakdowns}
Revisiting the conversational data through the lens of breakdowns and repair work allowed us to identify the diverse ways in which older adults navigated and patched conversations when they deviated from expectations, often requiring significant cognitive effort. These hidden, labor-intensive efforts underscored older adults’ adaptability, resourcefulness, as well as their recognition of the AI’s value. While their repair attempts were not always successful due to the rigid conversational structure of commercial AI systems, analyzing them nonetheless provided us valuable insights into the expectations that shaped these repair attempts. Specifically, they revealed the trial-and-error strategies and the external information sources in users’ environment that they envisioned as integral to improving AI interactions. These repair efforts were often improvisational rather than premeditated, reflecting situated actions rooted in participants’ contextual knowledge. They drew upon their environment, medication routines, and social networks to bridge communication gaps and work toward mutual understanding with the AI. This dynamic process of collaboration and conversational patchwork offers valuable insights into how AI systems can better align with user expectations and support personalized interactions. Despite challenges, participants remained optimistic about learning to interact effectively with the AI, often hesitating to place full blame on the system. As CG3 noted: \textit{“I do think we can make it work…we just need to know the right words and timings”}. Similarly, CG5 reflected: \textit{“Maybe I just need to know what it knows about me and what it does not…like does it know about other things in the house…our phones? calendars? Or even general details about medication as a task?”} These sentiments highlight both their willingness to adapt and the need for greater transparency, especially regarding the AI’s knowledge and reasoning, also reflected in the breakdowns in explanations when asked to provide justification for its utterances.

Additionally, we also found that this repair work often fell to caregivers, adding to their existing care work. CG4 reflected: \textit{“Honestly, while I like having it around and do think it could be useful, sometimes it feels like hard work trying to talk to her, and it feels like another thing I have to manage”}. Caregivers frequently stepped in to scaffold AI interactions for their loved ones, further illustrating the layered and collaborative nature of these efforts. While the AI functioned well as a basic check-in tool, we see that as conversations became more complex and personalized, the static nature of the AI within this human-AI collaboration increasingly led to participants experiencing frustration. CG2 described this “falling apart” during a follow-up interview: \textit{“You know, as long as she was asking us about medicines and we were saying yes we took it or we did not take it, it was okay. But as soon as I ask her to do something else, like remind me later, or ask her whether [husband’s name] has taken his morning meds, she just kind of falls apart…”}. This uneven burden of repair placed on users can significantly impact their long-term relationship with the AI, sometimes leading to interaction abandonment. We explore the implications of this imbalance in further depth in the discussion section.

\subsubsection{Speaking like a Human: An Inherent Anthropomorphic Expectation}

Our findings revealed that older adults tend to anthropomorphize Conversational AI assistants significantly. In our study, this anthropomorphization varied based on how participants perceived and attributed characteristics to the AI’s personality, often influenced by breakdowns in language and semantics. However, the expectation for AI systems to engage in human-like conversations was consistent across participants. This anthropomorphization was further evident in participants’ habitual references to the Google Home as “she” rather than “it”,  reinforcing the perception of interacting with a human-like entity. The voice-based modality, personalized greetings, and name usage further strengthened this effect. Additionally, expectations for confidence and explanatory ability were especially pronounced in high-stakes tasks like medication management. M1 expressed: \textit{“If she’s gonna tell me to take my medication, she better sound confident and assertive…like [my wife]},” underscoring the role of familiarity and assurance in such routines. These anthropomorphic expectations also emerged through the metaphors that participants used to describe the AI, likening it to a \textit{“friend”} (CG3), \textit{“teacher”} (CG2), \textit{“like my wife”} (M1), and even a familiar reference, \textit{“Siri”} (M4). Some extended these comparisons to fictional characters, such as Rosey the Robot from The Jetsons, envisioning a system that could become \textit{“like a part of our family and understand our needs and values”} (CG1)

Finally, we found that expectations for human-like interactions extended to conversational dynamics, particularly initiation, flow, and turn-taking. Most commercial AI assistants follow a “don’t speak until spoken to” model, driven by privacy policies. However, this places the burden of initiation solely on the user. In our study, while the AI initiated medication reminders at set times, it relied on users to adapt to its turn-taking protocols and conversational flow once the interaction began, that is generally expected in natural human conversations. Our interviews revealed that preferences for initiation vary based on context and the degree of anthropomorphism attributed to the AI, making \textit{one-size-fits-all} rules for initiation insufficient. Additionally, the rigidity of the AI’s conversational structure worsened breakdowns by limiting its ability to recognize, acknowledge, and integrate spontaneous user responses. In such a highly anthropomorphized interaction context, older adults are unlikely to form a distinct mental model of interacting with a technical AI system. Instead, their expectations, shaped by human-like metaphors and behaviors, highlight the need for systems that better align with human conversational norms. While we do not advocate for AI to position itself as a human interlocutor, an approach that raises ethical concerns and risks fostering overtrust, we do argue for embedding conversational nuances such as adaptable initiation, flow, and explanation into their design, without creating additional privacy harms.

\section{Discussion and Broader Implications}

\subsection{Learning from Human Conversational Strategies}

Designing effective Conversational AI requires balancing social and functional elements of human conversations that naturally incorporate strategies like turn-taking, topic maintenance, appropriate language use, and repair mechanisms to restore understanding.  Conversational repair, in particular, involves reestablishing mutual understanding by modifying or clarifying a message when it is not initially understood. Our study revealed that AI system often lack robust error recovery pathways, particularly for older adults. For example, when participants needed to correct medication records, they were unable to update the AI after an initial response due to its lack of historical memory and contextual awareness. To address this gap, error recovery pathways in advanced AI systems must be embedded into initial system models, anticipating a broad range of potential errors and assigning specific labels to interactions. This approach would allow users to revisit and amend relevant previous interactions, fostering a more flexible and adaptive conversational experience.

Furthermore, to contextualize the expectation for AI assistants to engage in human-like conversations, we draw on the Computers are Social Actors (CASA) paradigm introduced by Nass and Reeves \cite{reeves1996media}. CASA suggests that people unconsciously treat computers and digital agents as social entities, applying norms, expectations, and behaviors typically reserved for human interactions \cite{nass1994computers}. Similarly, Pradhan et al. found that older adults naturally assigned gender and polite cues to AI assistants \cite{pradhan2019phantom}, while Lopatovska et al. observed users frequently saying “thank you” and “please” in AI interactions \cite{lopatovska2019talk}. While this highlights the inherently social nature of conversational AI, it is important to acknowledge that expectations for natural, human-like conversations with AI systems remain debated. Clark et al. argue that human-machine interactions do not fully align with human-human conversations and should be treated as a distinct conversational genre \cite{clark2019makes}. However, building on the evidence analyzed and presented in our study, we see that user expectations are not limited to the system's ability to mimic human-like behavior or follow conversational conventions (ex: politeness, tone, responsiveness). Participants in our study also emphasized the practical and functional needs, including reliable assistance, adaptability to individual preferences, proactive explanations, and reduced interaction effort. This expectation is particularly pronounced in informal caregiving contexts, such as where AI systems support older adults at home \cite{desai2023metaphors}.

\subsection{Identifying Information Sources for Explanations in AI Systems}

In our study, we found a gap in the AI’s ability to explain or justify its responses to users in a way that is tailored to their expectations and goals. This breakdown further highlights the need for building more context-aware and human-centered explainable AI (XAI) systems. Our findings on the distributed nature of information in older adults’ environments contribute to this discourse, emphasizing the need for AI systems to look beyond their inner algorithmic processes for providing explanations. One approach (among others) is integrating external information sources within the user's sociotechnical context to provide more actionable and human-centered explanations \cite{mathur2024categorizing}. Alizadeh et al characterize this expectation for functional explanations as part of the AI considering a holistic user experience, rather than focusing on single incident queries, considering all relevant information sources \cite{alizadeh2023user, mathur2024categorizing}. In a longitudinal analysis, Pradhan highlights this interconnected network of human and non-human actors in older adults’ environment, such as digital and physical artifacts, rules, and routines, that shape their interactions with technology and argue for leveraging these connections in AI interactions \cite{pradhan2022more}. By bringing in these diverse information sources, AI can generate contextually situated explanations, also aligning with Ehsan et al’s argument that no single explanation is inherently sufficient and an explanation’s inherent value depends on how well it meets a user’s immediate goals, needs, and expectations \cite{ehsan2022human, ehsan2021operationalizing}.

In recent years, there have been growing calls for shifting the examination of explainability to a human-centered approach. This includes focusing on aspects of an AI’s decision-making processes that are relevant to the users’ sociotechnical context, rather than providing an overly technical description of the inner workings of the AI \cite{ehsan2020human}. As a result of this misalignment, a gap between user needs and the actual experience occurs and exacerbates existing issues with algorithmic opacity, or the “black-box” nature of current AI systems. While efforts have been made to categorize information for explanations \cite{arya2019one, wang2019designing, guidotti2018survey}, many lack specificity regarding the contextual understanding needed for assistive AI systems operating in environments like the home. Developing a robust framework for AI explanation sources requires research on phenotyping the diverse sociotechnical environments in which these systems operate, reflected in ongoing work exploring user perceptions of AI explanations in distinct contexts \cite{mathur2025research, kim2023help, mansi2023don}. Overall, we emphasize the importance of exploring the user’s sociotechnical context and engaging users directly in the design process, making explanation a shared and iterative task between AI and users. To the technical XAI community, we offer this insight: the critical input features AI systems use to make decisions or predictions are embedded within the broader sociotechnical environment. Yet, most XAI approaches assign weights to these features based solely on internal data-generated validation scores. We argue that identifying and weighting input features in XAI algorithms should follow a human-centered approach, one that actively involves end users and stakeholders in determining the most valuable information sources for generating meaningful AI explanations.

\subsection{Implications for Generative AI Systems} 

Given the rapid advancement in Generative AI-powered Large Language Models (LLMs), it is essential to reflect on our findings on conversational breakdowns and assess their relevance in the context of LLMs.  Specifically, we must examine whether these advanced systems can effectively mitigate the types of breakdowns documented in this work and, conversely, whether they introduce new and unforeseen breakdowns. A common assumption is that Generative AI’s sophisticated language abilities will inherently resolve conversational breakdowns. However, our findings indicate that misalignments are an inevitable aspect of human-AI interaction, even with state-of-the-art generative models. With techniques and frameworks such as Retrieval Augmented Generation (RAG) that can dynamically retrieve information from external sources (like databases, documents, or APIs) and use it to generate more informed and relevant responses, these systems can store and retrieve distributed information and recall conversational history. However, our findings indicate that effective conversational repair requires more than just recall; it depends on recognizing users’ intent  and being abie to interpret users’ immediate goals, contextual cues, and emotional signals to retrieve the most relevant information from user history. Without this nuanced understanding of intent, AI systems risk overwhelming users with irrelevant details or failing to provide responses that align with their expectations (what our participants described as “knowing me well”). This highlights the need for LLM-based AI systems to integrate not just memory and retrieval but also mechanisms for intent recognition and context-aware information retrieval to ensure conversational breakdowns are addressed in a natural and supportive manner.

Additionally, emerging research into the usability of LLMs in sensitive conversational contexts has also placed them under scrutiny for inducing hallucinatory responses, providing non-deterministic answers and exhibiting a lack of intent recognition. Shananan et al argue that LLM-powered chatbots in such contexts essentially become role-playing devices that can assume personas based on a user’s context \cite{shanahan2023role}. However, this has the potential to create misalignments, what Ferrario et al address as the risk of “social misattribution” by users towards LLM-powered chatbots in sensitive contexts such as mental health support \cite{ferrario2024addressing}. LLMs are often at a higher risk of such social misattribution due to their tendency to recall past user details and conversations, lack of accountability in case of conversational breakdowns \cite{quttainah2024cost}, and generating overtrust in their capabilities through expressions of uncertainty \cite{kim2024m}. Finally, in the context of explanations, we see that many of the concerns surrounding the opaque and black-box nature of traditional AI systems persist with LLMs. Balayn et al., through their user study, highlight that LLM-generated explanations still fall short in developing an accurate understanding of the sociotechnical context and the specific information needs of users \cite{balayn2024understanding}. This underscores an ongoing challenge of tailoring conversations in AI systems to align with the unique requirements and contexts of older adults despite advancements in AI capabilities.

\subsection{Broader Lessons for Human-AI Systems}
Here, we reflect on the broader implications of our findings, organized into three key points. \textit{First}, we discuss human-AI teaming as an effective approach for designing assistive Conversational AI systems for older adults, emphasizing collaboration where the AI and users complement each other’s strengths to address evolving needs. \textit{Second}, we reflect on the value of using the language of seams to reveal the often overlooked human effort involved in interacting with AI. By examining how users creatively adapt to these systems, we identify crucial information sources for AI and advocate for a strengths-based design approach that learns from users’ strategies to improve system design. \textit{Finally}, we end with briefly discussing user context through the lens of Critical Technical Practice (CTP), which uncovers hidden challenges and frictions in human-AI interactions. 

Human-AI teaming emphasizes that for AI systems to effectively establish common goals, they must collaborate with and learn from humans in ways that leverage human strengths and existing infrastructure \cite{berretta2023defining}. This approach fosters a partnership with AI systems rather than a tool-user relationship, which is crucial for building trust among older adults, a group often characterized as skeptical of technology in HCI research \cite{mcmurray2017importance}. In their investigation into older adults’ perspectives on adopting technological systems, Pena et al. found that older adults hesitate to adopt new technology primarily due to concerns about losing autonomy \cite{barros2021circumspect}. Therefore, it is essential that AI systems support users as partners rather than replacing tasks that they may value or find meaningful, such as caring for a partner or parent. Similarly, Yang et al. highlight human-AI interaction as a uniquely difficult design paradigm, where achieving a collaborative dialog flow is critical for human-centered AI systems \cite{yang2020re}. Our findings align with this, showing that AI systems are often conversationally unpredictable due to a lack of transparency and explainability of their inner mechanisms. While AI is typically designed to minimize human effort, we find that this perceived loss of autonomy can deter willingness to engage with it. Instead, users prefer AI as a compensatory support system that enhances, rather than replaces, their abilities. Teaming in this way has also been shown to improve task efficiency, with studies demonstrating that human-AI collaboration outperforms either human or AI working alone \cite{vaccaro2024combinations, bouschery2024artificial}.

To highlight how older adults navigated their interactions with the AI system in their homes, we applied Janet Vertesi’s interpretation of technological seams \cite{vertesi2014seamful}. The concept of seams in interactions serves as an analytical lens to reveal the \textit{ad hoc} efforts users make to negotiate with technological infrastructures to meet their information needs. This perspective helped us to focus on the work that participants did to make the AI system function effectively for them. Research in fields of data work has similarly examined human mediation in technology interactions, showing how this understanding can inform valuable design opportunities within sociotechnical environments \cite{wong2019work, dombrowski2012labor, ehrlich1999invisible}. By focusing on user-initiated repair, we gain insight into the immediate needs and resources that matter to users. This is often difficult to capture in structured lab settings or interviews as these repairs occur impromptu and organically in the real-world. Researchers may not always be able to observe how users react in the moment or address breakdowns, making in-context discovery a more authentic way to identify information sources that are of value to users in interactions with the AI system. 

This argument also supports adopting the reflective lens of Critical Technical Practice (CTP) in studying human-AI interactions. Introduced by Phil Agre in \textit{Computation and Human Experience} (1997), CTP critiques existing technical narratives that often obscure alternative perspectives and marginalize voices such as those of non-technical users in technology discussions \cite{agre1997computation}. Agre argues that addressing these blind spots requires first recognizing the dominant assumptions and metaphors within a field and then actively reframing them by identifying disruptions. In our work, studying conversational breakdowns experienced by older adults offers an alternative perspective on human-AI interaction. Examining the repair attempts performed by older adults revealed their needs and expectations in ways that highlighted the limitations of AI systems to handle conversational breakdowns. Instead of pursuing seamless and error-free interactions that are often designed with younger and more tech-friendly users in mind, our work advocates for a model that embraces breakdowns as opportunities to reflect on and improve conversational experiences, particularly for populations with diverse technological needs and expectations such as older adults. 

\section{Limitations and Future Work}

While this study offers important insights into conversational breakdowns and the identification of relevant information sources, a few limitations must be acknowledged. First, our findings are drawn from a participant sample already familiar with the AI system having engaged with it for non-medication tasks over a period of time, influencing their willingness to repair breakdowns. Interaction approaches for participants with no prior exposure to the AI remains an open area for investigation and may yield different patterns of breakdown repair. Second, this study was conducted within a specific sociotechnical context, older adults in a home environment, shaping the identified information sources accordingly. While we posit that these sources provide a foundational framework, their direct transferability to other contexts will require careful adaptation. In future work, we aim to refine and formalize this framework to enhance its generalizability. Additionally, in our future work, our goal is to leverage this framework to generate and evaluate AI-driven interactions, in particular, AI explanations, across diverse household scenarios, with the goal of understanding which types of explanations are most preferred in varying contexts of use.

\section{Conclusion}

As older adults continue to engage with AI systems in their homes in increasingly interpersonal ways, it is crucial to design such systems in ways that can robustly handle conversational breakdowns and provide error recovery pathways to mitigate those breakdowns. In this study, we investigated the longitudinal use of a Conversational AI system designed to support older adults’ medication management routines through an in-home deployment. By analyzing interaction data collected over the deployment period, we identified four primary categories of conversational breakdowns between older adult participants and the AI system. We observed that following all the breakdown instances, older adults attempted to repair or realign the conversation with the AI using their situated knowledge of their environment. Drawing on insights from analyzing these repair attempts, and through user interviews conducted with participants, we highlighted the need for AI systems to leverage the distributed knowledge in a user’s environment to enable more contextually aware interactions with them. We also reflected on this need for external information integration in light of AI explanations, and argue for AI systems to move beyond explaining their inner algorithmic processes to providing more understandable explanations to users.

\bibliographystyle{ACM-Reference-Format}
\bibliography{bibliography}


\end{document}